\documentclass[twocolumn,final,prb,superbib,byrevtex,superscriptaddress,showpacs,endnote]{revtex4}
\usepackage{setstack}
\usepackage{shortcuts}
\usepackage{amsfonts}
\usepackage{amssymb}
\usepackage{epsf}
\usepackage{psfrag}
\usepackage[pagewise,running]{lineno}
\usepackage{graphicx}
\usepackage[squaren]{SIunits}
\def\drts{{\textsc{drt}}}
\def\drt{{\textsc{drt}}}
\def\mc{{\textsc{mc}}}
\def\clsq{{\textsc{c-lsq}}}
\def\cnlsq{{\textsc{cnlsq}}}
\def\lsq{{\textsc{lsq}}}
\def\lfd{{\textsc{lfd}}}
\def\dc{{\textsc{dc}}}
\def\ppg{{\textsc{ppg}}}

\def\dd{{\text{d}}}
\def\ie{{\em{i.e.}}}

\def\rmd{{\rm{d}}}

\def\ve{{\varepsilon}}
\def\dashedd{{\,--\,--\,--\,}}

\begin{document}
\title{Resolving distribution of relaxation times in Poly(propylene glycol) on the crossover region}

\author{Enis Tuncer}
\email{enis.tuncer@physics.org}
\thanks{To whom correspondence should be addressed.}
\affiliation{Applied Condensed-Matter Physics, Department of Physics, University of Potsdam, Am Neuen Palais 10, D-14469 Potsdam Germany}
\author{Maurizio Furlani}
\affiliation{Solid State Physics, The {\AA}ngstr{\"o}m Laboratory, University of Uppsala, SE-751\ 21 Uppsala, Sweden}
\affiliation{Physics and Engineering Physics, Chalmers University of Technology and G{\"o}teborg University, SE-412\ 96 G{\"o}teborg Sweden}
\author{Bengt-Erik Mellander}
\affiliation{Physics and Engineering Physics, Chalmers University of Technology and G{\"o}teborg University, SE-412\ 96 G{\"o}teborg Sweden}
\date{\today}
\received{}
\accepted{}
\published{}

\begin{abstract}
In this paper, a recently developed numerical technique [{\em Tuncer~E and Guba{\'n}ski~S~M, IEEE\ Trans\ Diel\ El\ Insul\ {\bf 8}(3)(2001) 310-320}] is applied to poly(propylene glycol) complex dielectric data to extract more information about the molecular relaxation processes. The method is based on a constrained-least-squares (\clsq) data fitting procedure together with the Monte Carlo (\mc) method. We preselect the number of relaxation times with no {\em a-priori} physical assumption, and use the Debye single relaxation as ``kernel'', then the obtained weighting factors at each \mc\ step from the \clsq\ method builds up a relaxation time spectrum. When the analysis is repeated for data at different temperatures a {\em relaxation-image} is created. The obtained relaxation are analyzed using the Lorentz (Cauchy) distribution, which is a special form of the L{\'e}vy statistics. In the present report the $\beta$ and $\alpha$ relaxations are resolved for the \ppg. A comparison of the relaxations to those earlier reported in the literature indicate that the presented method provides additional information compared to methods based on empirical formulas. The distribution of relaxation times analysis is especially useful to probe the crossover region where the $\alpha$- and $\beta$- relaxations merge and the results show that the relaxation after the crossover region at higher temperatures is  Arrhenius-type as the $\beta$-relaxation. Moreover, this relaxation is more likely to be the continuation of the  $\beta$-relaxation, but with a different activation energy. %is more probable  was also applied to the present data. The \drt\ ananlysis probes to the crossover region, which is an indication of the intrinsic complexity of the glass transistion.
%the original data by single Debye relaxations resulting relaxation times ?(Di) and their amplitudes ?&epsiv;i,, yield the relaxation time spectrum, where i is equal or less than the number of data points. Two different predistributions of relaxation times are considered, log-uniform and adaptive. The adaptive predistribution is determined by the real part of the dielectric susceptibility ??', and it allows for the increase of the number of effective relaxation times used in the fitting procedure. Furthermore, since the number of unknowns is limited to the number of data points, the Monte Carlo technique is introduced. In this way, the fitting procedure is repeated many times with randomly selected relaxation times, and the number of relaxation times treated in the procedure becomes continuous. The proposed method is tested for `ideal' and measured data. Finally, the method is compared with a nonlinear curve fitting by a spectral function which consists of three contributions, i.e. the Havriliak-Negami relaxation polarization, low frequency dispersion and de conductivity. It has been found that more information can be obtained from a particular data set if it is compared with a nonlinear curve fitting procedure. The method also can be used instead of the Kramers-Kronig transformation 
\end{abstract}


\maketitle
\section{Introduction}
%Information about the electrical properties of materials for specific application can be obtained by isothermal~\cite{Jonscher1983,MacDonald1987} or thermally stimulated  measurements~\cite{Turnhout1972}. In the former case, dielectric spectroscopy and time domain measurement techniques have been widely used. Although, only an analysis technique for dielectric spectroscopy data is presented and discussed here, the same procedure can as well be  applied to isothermal transient and thermally stimulated current data. % to obtain the relaxation maps~\cite{Tuncer1998a,Bello1995}.

Traditional methods for analyzing dielectric data as well as impedance data consist of applying curve-fitting-algorithms using empirical formulas [see for example \textcite{MacDonald1987}]. By such analysis information on  conductivity, molecular relaxation, liquid-glass transition, {\em etc.}, of materials are obtained which are challenging subjects in condensed matter physics. 
%In those approaches the fitting parameters are plotted against state variables such as temperature, pressure, electric field, {\em etc.} 
A possible new overture that has previously been avoided by researchers and instrument manufacturers has been the distribution of relaxation times (\drts) approach. Recently a computation method has been developed by \textcite{Tuncer2000b} that can resolve unique \drts\ for dielectric data, see for example \textcite{Tuncer2000b}, \textcite{Tuncer2002b} and \textcite{Tuncer2002elec}. The method is based on the Monte Carlo (\mc) technique \cite{Binder} and the constrained-linear-least-squares (\clsq) algorithm~\cite{Adlers}. The procedure obeys the Kramer-Kronig relations \cite{LL}. It is, therefore, distinct to acquire ohmic-losses even if their contributions are not visible to the naked eye. Finally, once the \drts\ is known it is trivial to obtain the time-domain relaxation function--{\em the responce function}.
\begin{figure}[t]
  %\newpage
  \begin{center}
    {\includegraphics[width=1.2in]{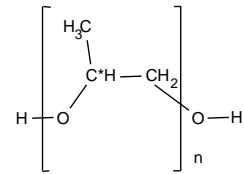}}
  \end{center}
  \caption{\label{fig:structureppg} Chemical structure of poly(propylene glycol) with the average molecular weight of $4000$, the number of structural units are approx. $69$. The asterisk ($*$) indicates the chiral carbon. In atactic \ppg\ the configuration (R or S) of the chiral carbon is randomly distributed  along the chain.}
\end{figure}

\begin{figure}[t]
  \centering
  {\includegraphics[width=2.3in]{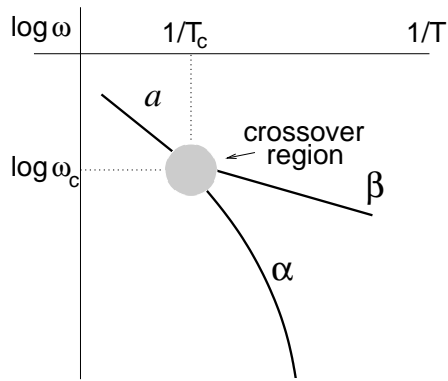}}
  \caption{Crossover region in the Arrhenius diagram~\cite{GlassTrans}. $a$ is the high temperature process, $\beta$ the local Arrhenius process (Johari-Goldstein process), and $\alpha$ the cooperative process.}
  \label{fig:crossover}
\end{figure}

%Information about molecular relaxation processes  and the morphology of the considered system can be obtained using various analyses techniques. 
None of the conventional techniques can resolve dipolar dielectric relaxations from measured data that contain contributions of several  processes and because of measurement limitations. So in this paper, the developed numerical method \cite{Tuncer2000b} is applied to analyze dielectric data of atactic poly(propylene glycol) (\ppg) (see Fig.~\ref{fig:structureppg}).  This polymer has a branched structure which is different from the more commonly used poly(ethylene glycol). A random distribution of the absolute configuration of the tertiary carbon leads to a large number of different sequences, and therefore, to a complete lack of crystallinity also at the lowest temperatures. Anyhow, the orientation of the structural unit in the chain, quite regularly head to tail, leads to a cumulative dipole moment along the chain contour~\cite{Mijovic2002} (Stockmayer type-A polymer). In \ppg, %presence of ({\em i}) local dipoles correspondent to the oxygen atoms, ({\em ii}) global dipoles along the chain \cite{Stockmayer}, and ({\em iii}) the relaxation of methyl-groups \cite{Johari} give rise to 
 several relaxation phenomena are observable in the dielectric spectroscopy \cite{Johari,Varadarajan,FKremerPPG,Gomez2001,Schroter1998,Leon1999}. The relaxations are labeled as $\alpha'$, $\alpha$ and $\beta$ in dielectric data. The $\alpha'$-relaxation, also known as the normal-mode relaxation, is strongle dependent on molecular weight, and its amplitude is lower than the other ones. The $\alpha$-relaxation can be related to the cooperative movements of the segments of the macromolecular chains and its temperature dependence. The $\beta$-relaxation is suggested to be associated with localized motions of the structural units and it has a weaker Arrhenius-type temperature dependence\cite{AngellNgai,DixonPRL}. 
Recent studies of dielectric relaxation in glass-forming liquids have shown a complex phenomenon, which is known as the crossover region \cite{GlassTrans,Donth2002,Schroter1998,AngellNgai,Beiner2001}. In this region, $\alpha$- and $\beta$-relaxations merge in the Arrhenius plot (see Fig.~\ref{fig:crossover}), after the merging point, one relaxation survives whose origin is sought. The crossover region in glass forming liquids can in general be described as a transition window from a molecular liquid at high temperatures ($a$ process) to a cold liquid where cooperative process ($\alpha$ process) is present \cite{GlassTrans} as presented in Figure~\ref{fig:crossover} where $T_c$ and $\log \omega_c$ are temperature and angular frequency at the crossover. In other words at high temperature, above the crossover region, one can observe the molecular characteristic relaxations, \ie\ the local rearrangements, while below that region one can see two different processes; associated with local and %the enhancement of the macro molecular and 
collective processes. 
%where the polymer characteristics are enhanced respect to the local molecular characteristics~\cite{GlassTrans} in the cold liquid, as presented in Figure~\ref{fig:crossover} with $T_c$ and $\log\omega_c$. 
We therefore focus on the crossover region in \ppg, and probe this region with the \drt\ analysis in this study. %It is therefore possible to calculate different activation energies for each process. 
%{\em to establish its strenght}.
\section{Background on dielectric data analysis}

When dielectric data are obtained over a wide frequency range, not only contributions of relaxation polarization, direct-current (\dc) conduction and low frequency dispersion (\lfd) but also contributions of space charges, electrode effects {\em etc.}, may be present. If only the contributions of relaxation polarization, \dc\ conduction and \lfd\ are taken into account, it is possible to distinguish the \dc\ conduction among the other processes. Kramers-Kronig relations \cite{LL} are often used for this intention. However this requires that, ({\em i}) the dielectric system gives a linear response and ({\em ii}) the frequency range should be wide enough for the integration. The second condition is not usually possible for most experiments and equipment. There are several other ways to separate the \dc\ conductivity contribution such as using pairs of Debye functions or Curie von Schweidler power-law functions--which are in principle the same as Kramer-Kronig transformations and finally performing measurements in the time domain \cite{Gafvert1996,Jonscher1983}. %In addition to these techniques, complex nonlinear least squares analyses could be applied~\cite{MacDonald1987}. %A more general way of dielectric 

Besides, data analyses can be carried out in the frequency domain as well as in the time domain using, for example, complex non-linear-least-squares (\cnlsq) \cite{MacDonald1987,Snyder1998,Marquardt1963,Tsai1982,MacDonald1986,Tuncer2001a} or \drts\ methods~\cite{Tuncer2000b,Keiter98,Kliem88,Tittelbach93,Carmona99}. In the former technique, an analytical expression of the model is needed.  The expression can be in the form of an equivalent circuit \cite{MacDonald1987} or of known relaxation functions \cite{Al-Refaie1996,Snyder1998} or even as an addition of several functions \cite{HillLFD,Capaccioli1998,Tuncer2000a}.  However, \textcite{Factor1996} has stated that for a narrow range of data \cnlsq\ is not reliable. 

When there is no analytical model avaliable for the relaxation behavior or single relaxation processes are of importance \drts\ can be preferred. Approaches with known-distribution-functions for \drts\ have been proposed in the past~\cite{Bottcher,Daniel,Turnhout1972}. The most general ones are either Havrialiak-Negami~\cite{HN} relaxation polarization or its derivatives~\cite{Jonscher1983}; Cole-Cole~\cite{CC} and Cole-Davidson~\cite{CD}. There have been several publications on the physical bases of these analytical functions \cite{Kirkwood1941,Ngai1979,Schlonhals1989,Tuncer2002b}. There have also been a couple of critical comments in the literature regarding the \drts\ and its meaning, \ie, %here  have been couple of comments in the literature i.e.,  %~\cite{KirkFuoss,jons90}
%   \textcite{KirkFuoss} and \textcite{jons90} have stated %, Fuoss and Kirkwood J. Am. Chem. Soc. {\bf 63}, 385 (1941).} 
%\begin{quotation}
    {{\em ``...We have, however, no actual proof of necessity for the existence of a distribution of relaxation times for a given system, nor have we any right to assume that a possible distribution should be Gaussian''}}~\cite{KirkFuoss}
%  \end{quotation}
%  \begin{quotation}
and    {{\em ``...such theories as distribution of relaxation times which are purely phenomenological and are not based on any rigorous physical arguments. There are also empirical mathematical functions which represent the experimentally observed results to a certain extent, but which are not in any way related to the physical nature of the phenomena in question''}} \cite{jons90}. %Jonscher, IEEE Trans. on El. Insul.  {\bf 25}(4), 622 (1990).}
%  \end{quotation}
When more relaxations coincide in energy or they overlap in part, researchers try to choose appropriate compositions, temperature windows and molecular weights to avoid the coincidence, but this is not always possible. The analytical functions~\cite{Gomez2001,WilliamsAnsatz,WilliamsWatts}, on the other hand, are not often  capable of resolving overlapping relaxations \footnote{See for example Ref. \onlinecite{Leon1999}, which discusses the absence of $\beta$-relaxation in propylene glycol and its being  masked by the $\alpha$-relaxation. The proposed method can be applied to such systems without difficulty}. The mentioned functions are extensively used to obtain average or most probable relaxation times. As a result, the \drts\ have  been more or less avoided as an established dielectric data analysis tool, though there have been couple of numerical attempts \cite{Kliem88,Tittelbach93,Carmona99,Morgan}. A mathematical method to resolve the \drts\ has also been proposed by \textcite{Kremer} and \textcite{Winterhalter} who used Tikholov regularization algorithms. \cnlsq\ has also been applied to resolve the \drts~\cite{Macdonald1995,Macdonald2000a,Macdonald2000b,Mustapha2000}. In addition arbitrary analytical distributions were also proposed and used in the \cnlsq\ algorithms~\cite{Al-Refaie1996,Gomez2001,Schroter1998}. As they mention, this ill-posed problem can not be solved using conventional \lsq\ and linear programming but in this present paper we show  {\em the usefulness and power of the \mc\ technique} for this purpose. In the next section we briefly review this issue in which we combine the \clsq\ algorithm \cite{Adlers} with the \mc\ technique. %It is shown that one could obtain more information using the frequency dependence of the dielectric relaxation of materials by resolving the distribution of relaxation times.% (see \textcite{Tuncer2000b} for details).

\section{Resolving distribution of relaxation times}

\subsection{Numerical procedure}

The complex dielectric permittivity, $\varepsilon=\varepsilon'-\imath\varepsilon''$, of materials in the distribution of relaxation times approach is expressed as follows 
\begin{eqnarray}
\label{eq:2}
  \varepsilon(\omega)&=&\varepsilon'(\omega>{\omega}_\uparrow)\\\nonumber
  &&+\Delta\varepsilon\int_{-\infty}^{\infty}{{\sf g}(\log \tau)}[1+\imath\omega\tau]^{-1} \rmd \log \tau 
\end{eqnarray}
where $\omega$ is the angular frequency ($\omega=2\pi\nu$) of data in the frequency window with a lower bound $\omega_\downarrow$ and with a high bound $\omega_\uparrow$ ($\omega_\downarrow\le\omega\le\omega_\uparrow$). $\tau$ is the relaxation time of the process with the total dielectric strenght $\Delta\varepsilon$. $\varepsilon'(\omega>\omega_\uparrow)$ is the instantaneous permittivity value that contains all relaxations faster than the considered frequency window.  Our numerical procedure extracts the unknown distribution function ${\sf g}(\tau)$~\cite{McCrum} in Eq.~(\ref{eq:2}) from the dielectric spectrum without any {\em a-priori physical assumption}. The relaxation distribution function must satisfy the condition
\begin{equation}
    \int_{-\infty}^{\infty}{\sf g}(\log \tau) \rmd \log \tau = 1  \label{eq:1}
\end{equation}
Before continuing any further, attention should be paid to the frequency dependent permittivity $\varepsilon(\omega)$ in Eq.~(\ref{eq:2}) which involves an integral dependent on $\rmd \log \tau$. In numeric analysis, this is undefined (or ill-mannered) because one has a finite number of data points $N$ -- in this case we have to use discrete $\tau$ values such that the number of $\tau$'s ($N_0$) should not exceed $N$. So Eq.~(\ref{eq:2}) is instead written as a sum of individual relaxations.
\begin{eqnarray}
    \varepsilon(\omega)=&\varepsilon'(\omega>{\omega}_\uparrow) + \sum_i^{N_0} {\Delta\ve_i}[{ 1+\imath \omega \tau_{D_i}}]^{-1} 
    \label{eq:3}
\end{eqnarray}
In this approximation, the Maxwell superposition principle with non-interacting independent dipoles (assumption of \textcite{Debye1945}) with relaxation times $\tau_D$ and dielectric strengths $\Delta\varepsilon$ are  assumed.  Then using \cnlsq, one can find pairs of $[\tau_D,\Delta\varepsilon]$. However, the pairs would not be unique--we get different $[\tau_D,\Delta\varepsilon]$-sets for different initial guesses or in other words the solutions are not orthogonal to each other. Our innovation in the method comes here, we introduce a {\em pre-distribution} of relaxation times ($\tau$). Fixing the $N_0$ number of $\tau_D$-values, the optimization problem becomes linear and can be solved by \clsq--the only unknowns are $\Delta\varepsilon_i$-values in this case which satisfy
\begin{equation}
\Delta\varepsilon=\sum_i^{N_0} \Delta\varepsilon_i
  \label{eq:totaldielec}
\end{equation}
Simultaneously, the values of $\Delta\varepsilon_i$ should not be negative due to the physical insignificance of the results; we neglect any possible resonances in the dielectric spectrum. This leads to assigning constraints on $\Delta\varepsilon_i$, and the outcome of the least-squares for the $\Delta\varepsilon_i$ are all positive. Since there is a one-to-one relation between the pre-distributed $\tau_{D_i}$-values and the resulting dielectric strengths $\Delta\varepsilon_i$, the probability densities of relaxation times $\tau_{D_i}$ are
\begin{equation}
  \label{eq:spectrum}
  {\sf g}(\tau_{D_i})=\frac{ \Delta\varepsilon_i}{ \Delta\varepsilon}
\end{equation}
This is the resulting relaxation time spectrum, ${\sf g}(\tau_{D_i})$, and $\Delta\varepsilon$ is defined in Eq.~(\ref{eq:totaldielec}). However, keep in mind that a single optimization run for ${\sf g}(\tau)$ does not mean anything--the solution is not unique and the time-axis is discrete. To overcome this, a \mc\ procedure is introduced to obtain the $[\tau_D, \Delta\varepsilon]$ pairs--the optimization repeated many times ($n$)--for randomly selected $\tau_{D}$ values \cite{NRBook} using a pre-distribution function. For the calculations the time-space can be assumed to be continuous. The distribution of the generated $\tau_{D_i}$ and the obtained $\Delta\varepsilon_i$ from \clsq\ optimization determine the relaxation time spectrum, ${\sf g}(\log\tau_D)$.

The obtained distributions are analyzed by means of comparing them with known distributions. We apply the L{\'e}vy statistics~\cite{BreimanLevy,LoeveBook,Walter1999,Donth2002}, which is used for interacting systems in different research fields~\cite{Levy,Barkai2000,Barkai2002,Furukawa1993,Stoneham1969,Walter1999,Donth2002}. The L{\'e}vy stable distribution is a natural generalization (approximation) of the normal (Gaussian), Cauchy or Lorenz and Gamma distributions. It is used when analyzing sums of independent identically distributed random variables by a diverging variance. %The L{\'e}vy stable probability density replaces the normal distribution in those cases, and 
Its characteristic function is expressed as \endnote{Different forms of probablility density functions for L{\'e}vy statistics exists, we have adopted a stable distribution used in Refs.~\onlinecite{BreimanLevy,LoeveBook,Walter1999}. We omit the imaginary parts in the characteristic function because of their insignificance in the results.}
%\begin{widetext}
\begin{eqnarray}
  \label{eq:Levy}
  {\sf L}(x;A,\mu,\gamma,\zeta)=A|\exp\{ - |\zeta(x-\mu)|^\gamma\}|
\end{eqnarray}
%\end{widetext}
Here, $0<\gamma\le2$. We need only four parameters; $\gamma$ characteristic exponent, $\mu$ localization parameter, $\zeta$ scale parameter and $A$ amplitude. The special forms of Eq.~(\ref{eq:Levy}) are the Gaussian [${\sf L}(x;\,A,\,\mu,\,2,\,\zeta)$], the Lorentz or Cauchy [${\sf L}(x;\,A,\,\mu,\,1,\,\zeta)$] and Gamma [${\sf L}(x;\,A,\,\mu,\,1/2,\,\zeta)$] distributions. This equation could be used to characterize the properties of the obtained \drt\ with $A$ corresponding to the amplitute of the mean relaxation time--its dielectric strength.

\subsection{Illustrative examples}
\subsubsection{Delta function (distribution)}
\begin{figure}[t]
  %\newpage
  \begin{center}
    {\includegraphics[width=3in]{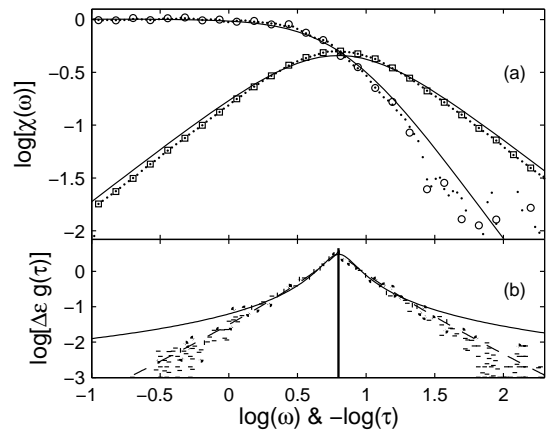}} 
  \end{center}
  \caption{\label{fig:debye} The Monte Carlo analysis of Debye relaxation with relaxation time $\tau=(2\pi)^{-1}\ \second ~(-\log(\tau)\sim0.798)$ and $\Delta\varepsilon=1$. (a) The real and imaginary parts of the dielectric susceptibility, $\chi=\varepsilon-\varepsilon'(\omega>{\omega}_\uparrow)$ and (b) the \drt calculated with the presented procedure. The smooth solid (\full) line in (b) is the fit obtained from the Lorentz (Cauchy) distribution with $n=8.165~(\text{FWHM}=0.245)$, $'log\bar{\tau}=-0.801$ and $A=0.018$. The dashed line (\dashedd) in (b) is the L{\'e}vy stable distribution, ${\sf L}(\log\tau;\,3.768,\,-0.801,\,0.834,\,\,8.167)$.}
\end{figure}

A distribution in the form of a delta function yields the integral in Eq.~(\ref{eq:2}) to be equal to the Debye relaxation function, ${\sf g}(\log \tau)=\delta(\log \tau-\log \tau_0)$ with the relaxation time $\tau_0$. We have added $1\% $ Gaussian random error to the data. The data is generated by assuming a relaxation with $\tau=(2\pi)^{-1}\ \second$ with $\Delta\varepsilon=1$. In Figure~\ref{fig:debye}, the results from the dielectric analyses are shown. In the analysis 32 relaxations are used and the Monte Carlo procedure is repeated for 6400 times. The solid (\full) lines in Figure~\ref{fig:debye}a are the fits obtained after the reconstruction of the original data. The \drt\ is presented in Figure~\ref{fig:debye}b is analyzed using Eq.~(\ref{eq:Levy}). The characteristic exponent yields the distribution is one of the special cases of Eq.~(\ref{eq:Levy}), Lorentz or Cauchy distribution, $\gamma\sim1$ and $\xi\sim0$. The Cauchy and the L{\'e}vy distributions are illustrated with the solid (\full) and dashed lines (\dashedd), respectively. Since the number of unknown parameters in the Cauchy is lower than the L{\'e}vy's, the Cauchy distribution is used in the following discussions regarding the \ppg\ data. Moreover, the Cauchy distribution is also one of the delta sequences ${\sf C}(\tau;A,\bar{\tau},n)$~\cite{Butkov}\endnote{\protect Other delta sequences are $\delta(x)=n/\sqrt{\pi}\exp(-n^2x^2)$ {and} $\delta(x)=1/(n\pi)\sin^2(nx)/x^2$.},
\begin{eqnarray}
  \label{eq:Cauchy}
  {\sf C}(\tau;A,\bar{\tau},n) &\simeq&  {\sf L}(x;A,\,\mu,\,1,\,\zeta) \\\nonumber
  &=&A\,n\pi^{-1}\{1+[n(\log\tau-\log\bar{\tau})]^2\}^{-1}
\end{eqnarray}
 where $A$ and $n$ are fitting parameters and $\bar{\tau}$ is the most probable relaxation time. The actual relaxation time is also displayed  at $-\log \tau=\log 2\pi=0.8$ as a vertical line in the figure. Eq.~(\ref{eq:Cauchy}) can also be converted to the Breit-Wigner (also known as the Lorentz) distribution, which is a general form originally introduced to describe the cross-section of resonant nuclear scattering, by substituting $n$ with $2/\Delta\tau$, where $\Delta\tau$ is the width at half maximum (FWHM). Although $\bar{\tau}$ and $\Delta\tau$ are related to each other with Heisenberg's uncentainity principle ($\bar{\tau}\Delta\tau=h/(2\pi)$; $h$ is Plack constant), we do not consider any quantum mechanical effects during the generation  of the data, and  there is no such relation between two parameters $\bar{\tau}$ and $\Delta\tau$. For various simulations of the Debye relaxation yields $6.5\le n\le10$ or $0.2\le\Delta\tau\le0.3$, the range is strongly dependent on the number of the Monte Carlo loops, and the actual position of the relaxation in the experimental window. We have also used Eq.~(\ref{eq:Levy}) to fit the results, this equation was more successful  for lower values of the \drt, at high values there is no difference between the general L{\'e}vy and the Cauchy distribution. %Finally, Eq.~(\ref{eq:4}) is also a special case of Levy distribution~\cite{GlassTrans}.
Summing up the obtained distribution is verified to be a delta sequence, ${\sf C}(\tau;A,\bar{\tau},n)=\delta(\tau-\tau_D)$

The comparison between the generated data and the extracted distribution illustrates a line broadening \cite{Stoneham1969}, which is due to several factors. First of all we have added some Gaussian noise to the data, although this is not as significant as the other factors, it influences the distribution. Secondly, the $\tau_D$ values are selected randomly from a log-linear distribution [see Ref. \cite{Tuncer2000b} for details]. If a biased distribution is used \ie\ based on the derivative of $\varepsilon'$ with respect to frequency ($\dd\log\varepsilon/\dd\log\omega$), the width of the distribution get narrower. However, it is nontrivial to find the initial distribution. If $\dd\log\varepsilon/\dd\log\omega$ is used the resulting distribution is not smooth enough due to numerical derivation problems. The numerical integration in Eq.~(\ref{eq:3}) is the other reason, which because of the calculation-speed considerations is based on a simple numerical summation with unit base of each $\tau_D$ generated. It can be improved by altering the integration routine. Finally, we can further improve the solution by increasing the \mc\ loops, which is a  statistical improvement. As a final remark, one can also argue the significance of a delta function in the concept of theory of distributions and real systems in the Nature \cite{Stoneham1969}.
\subsubsection{Box distribution}

\begin{figure}[t]
  %\newpage
  \begin{center}
    {\includegraphics[width=3in]{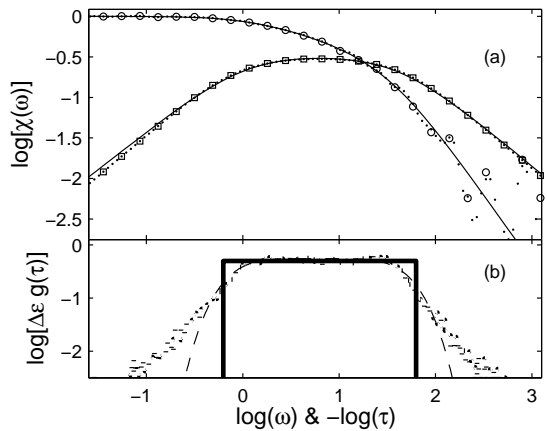}} 
  \end{center}
  \caption{\label{fig:box} The Monte Carlo analysis of the box distribution. (a) The real and imaginary parts of the dielectric susceptibility, $\chi=\varepsilon-\varepsilon'(\omega>{\omega}_\uparrow)$ and (b) the \drt\ calculated with the presented procedure. The thick solid (\full) line in (b) is the original box distribution. The dashed line (\dashedd) is ${\sf L}(\log\tau;\,0.553,\,-0.788,\,4.740,\,\,1.019)$.}
\end{figure}

Now, we apply the method presented to a box distribution. The box empirical distributions has been described in detail for mechanical relaxations~\cite{Tobolsky1960,McCrum} and in the Fr{\"o}chlich molecular theory of relaxation~\cite{Frohlich}. The box distribution is defined by the following equations
\begin{eqnarray}
  \label{eq:5}
  H(\log \tau) = \text{constant} \quad &\text{for} \qquad \tau_1\le\tau\le\tau_2\nonumber\\ 
  H(\log \tau) = 0 \quad &\text{for} \qquad \tau>\tau_2;\tau<\tau_1 
\end{eqnarray}
The results are presented in Figure~\ref{fig:box}a and \ref{fig:box}b, which show the reconstructed and obtained \drt, respectively. The original box distribution is also presented in Figure~\ref{fig:box}b as thick rectangle which indicate a good aggreement with the \drt. We have also applied Eq.~(\ref{eq:Levy}) to fit the \drt. The result is shown as a dashed line (\dashedd), ${\sf L}(\log\tau;\,0.553,\,-0.788,\,4.740,\,\,1.019)$. The generalized spectral line expression is successful to model even broad spectral lines as the box distribution.

\section{Experimental}
\ppg\ of weight-average molecular weight 4000 (purchased from Polysciences Inc., USA) was used after high vacuum drying at $10\ \micro\pascal$ at $80\ \celsius$ for $24\ \hour$. The cell was mounted in an inert atmosphere. The dielectric measurements on \ppg\ samples were performed at temperatures between $248\ \kelvin$ and $344\ \kelvin$ in an angular frequency $\omega$ range $2\pi\,10^6\ \reciprocal\second-3.6\pi\,10^9\ \reciprocal\second$, using a Hewlett Packard HP 4291 A HF impedance/material analyser together with a Novocontrol BDS 2100 sample cell and a Novocontrol 2200 RF extension line. The amplitude of the signal was $0.1\ \volt$.

\section{Application of the method}
Parameters used in the \mc\ analyzing procedure were as follows
  \begin{eqnarray}
      N>64 \quad N_0=32\quad n=10000 \nonumber
  \end{eqnarray}
The number of $\tau_D$-values were $3.2\times10^5$ in the calculations. Computations were performed on a Pentium $4$ ($2.4\ \giga\hertz$ {\tt LinuxPc}) with $1\ \giga\mathrm{byte}$ memory. Analyzing each dielectric data set takes between $3$ to $4\ \hour$ with the parameters selected above.

It is very important to assign an instantaneous permittivity value to start the analyses--relaxations over $\omega_\uparrow$ are summed in this permittivity value, here $\omega_\uparrow$ is the highest angular frequency used in the experiments. This is not actually just a constant number substracted from the real part of the dielectric permittivity since the data (the dielectric susceptibility) must afterwards satisfy the Kramer-Kronig relation. For this reason the procedure in \textcite{Nikl} has first been adapted, however, in some data sets this procedure has not produced good-fits at high frequencies.% and (ii) there are \drts\ at high frequencies which do not move with temperature; generates an undesired numerical-artifact. 
Therefore, a new preprocessing method was introduced in which the high frequency part of data was curve-fitted by a Havriliak-Negami empirical formula [Eq.~(\ref{eq:7})] \cite{HN} with a \cnlsq\ algorithm with a  fast relaxation time $
\tau\approx\omega_\uparrow$. Later, the data-set is extrapolated by this empirical formula, which significantly improves the \drts-spectra. Obtained \drts\ is analyzed by Eq.~(\ref{eq:Cauchy}) and the results are compared with those obtained from the curve-fitting analyses using the Havriliak-Negami expression,
\begin{equation}
  \label{eq:7}
  \varepsilon=\varepsilon'(\omega>{\omega}_\uparrow)+\Delta\varepsilon[1+(\imath\omega\tau)^\alpha]^{-\beta}
\end{equation}
where, $\Delta\varepsilon$, $\tau$, $\alpha$ and $\beta$ are the fitting parameters.
\section{Results and discussions}
\label{sec:results-discussions}
\begin{figure}[t]
  %\newpage
  \begin{center}
    {\includegraphics[]{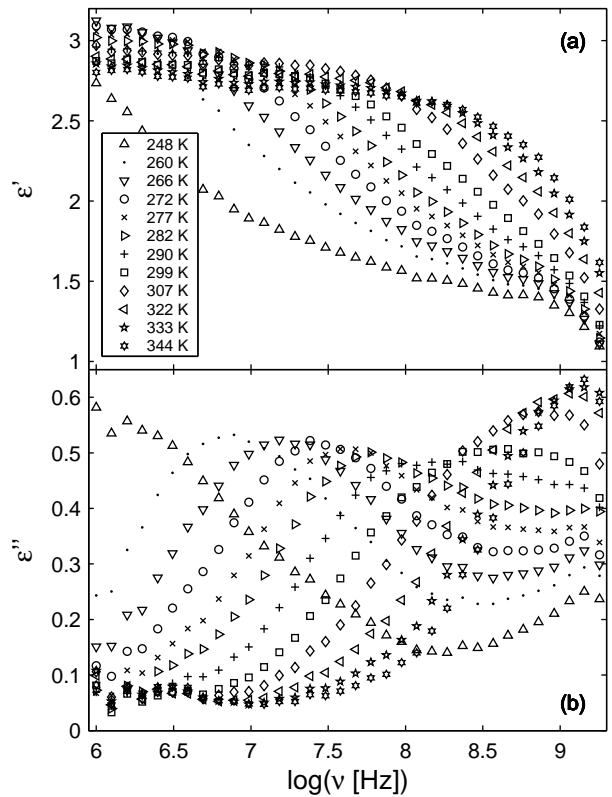}} 
  \end{center}
  \caption{\label{fig:alleps} (a)Real $\varepsilon'$ and (b) imaginary $\varepsilon''$ parts of permittivity versus logarithm frequency $\nu$ at various temperatures.}
\end{figure}

Figure~\ref{fig:alleps} shows the real and imaginary parts of the dielectric permittivity for \ppg\ over the avaliable frequency range. Each curve corresponds to a different temperature. The $\alpha$-relaxation is shown as a distinct peak, and the onset the $\beta$-relaxation can be observed at the lowest temperatures. At temperatures over $300\ \kelvin$ the two peaks overlap, producing an increase in the magnitude of the dielectric strength. At temperatures above $300\ \kelvin$ a weak $\alpha'$-relaxation may be seen at low frequencies. The frequency of the dominant $\alpha$-peak increases with temperature and its amplitude slightly decreases with increasing temperature.% the temperature is increased the peak frequency $\omega_p$ increases. The amplitude of the imaginary part of the permittivity first decreases and after $290\ \kelvin$ it increases. Its width broadens, and at high temperatures the shape of $\varepsilon''$ becomes very broad exhibiting no sign of relaxation peak in the experimental frequency window. 
This behavior is characteristics of many molecular systems, {i.e.}, see for example, \textcite{Furukawa} for \ppg\ and \textcite{DixonPRL}, \textcite{FKremerPRL} and \textcite{AngellNgai} for other glass-forming liquids. Although curve-fitting approaches using a Fourier transform of streched exponential \cite{Kohl,Phillips} or Havrialiak-Negami \cite{HN} as well as its derivatives would be an appropriate method for the analysis, it does not produce valuable information regarding individual molecular processes other than the most probable one \cite{AngellNgai}. In addition, it is worth mentioning that at temperatures higher than the glass-transition temperature, it is not easy to observe the $\beta$-relaxation in some molecular systems since it is hindered by the stronger $\alpha$-relaxation in the crossover region. Finally, the dielectric loss of \ppg\ does not  exhibit any sign of ohmic-losses at high temperatures and at low frequencies. 
\begin{figure}[t]
  %\newpage
  \begin{center}
    {\includegraphics[]{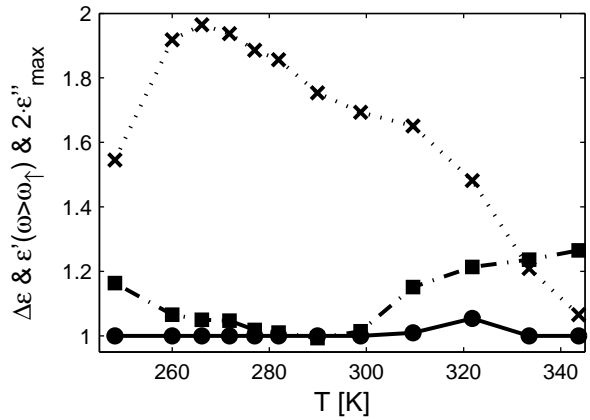}}%[width=3.3in]{eps0temp.eps}} 
  \end{center}
  \caption{\label{fig:eps0temp}Dielectric permittivity at high frequencies $\varepsilon'(\omega>\omega_\uparrow)$ ($\bullet$), double maximum dielectric loss $2\varepsilon''_{\max}$ ($\blacksquare$) and $\Delta\varepsilon$ ($\times$) as a function of temperature. Lines are drawn to guide the eye.}
\end{figure}

Figure~\ref{fig:eps0temp} shows the instantaneous dielectric permittivity as a function of absolute temperature at high frequencies $\varepsilon'(\omega>\omega_\uparrow)$, double maximum loss $2\cdot\varepsilon''_{\max}$ and calculated dielectric strength $\Delta\varepsilon$. The permittivity value fluctuates between $1$ and $1.05$ at temperatures lower than $300\ \kelvin$. It indicates that the sizes as well as the number of dipoles with fast relaxation times are small at frequencies higher than $1\ \giga\hertz$. At higher temperatures instantaneous permittivity increases--this means that a couple of relaxations are finalized and they contribute to instantaneous polarization. %The maximum of loss-peak $\varepsilon_{\max}$ decreases as temperature approaches to $300\ \kelvin$ and later starts to increase again. 
The estimated total dielectric strength $\Delta\varepsilon$, on the other hand, neglecting the first data point (the first data point could be due to virgin sample, which experiences the temperature and electric field the first time), decreases as the temperature increases. Taking into account the Debye relaxation model \cite{Debye1945,Frohlich}, which states that $\Delta\varepsilon\equiv 2\cdot\varepsilon''_{\max}$, the discrepancy between the calculated $\Delta\varepsilon$ and double the measured maximum loss-peak $2\cdot\varepsilon''_{\max}$ proves that the relaxations at temperature lower than $320\ \kelvin$ are broad (non-exponential decay function) and can be modeled by the \drts. The data also indicates that there can be several single relaxations with relaxation times close to each other. However, at temperatures higher than $320\ \kelvin$ these two values get close to each other indicating that the relaxation is nearly-exponential. The value of $2\cdot\varepsilon''_{\max}$ must not exceed the dielectric strength, but at the highest temperatures this condition is not satisfied, which is due to lack of enough data points at high frequencies--we have no $[\tau, \Delta\varepsilon]$ pairs at high frequencies from the calculations. 
%Finally, the behavior of $2\cdot\varepsilon''_{\max}$ and $\Delta\varepsilon$ change after $300\ \kelvin$ as the temperature is increased, the latter jumps to a higher value and increses slightly with temperature. $\dd\varepsilon''_{\max}/\dd T$ on the other hand becomes stepper higher temperature than this temperature. 

\begin{figure}[t]
  %\newpage
  \begin{center}
    {\includegraphics[]{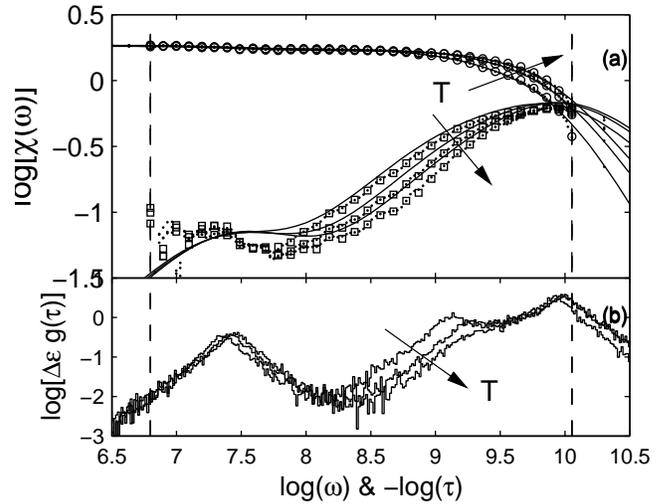}}
  \end{center}
  \caption{\label{fig:temp130} (a) Dielectric susceptibilities $\chi$ versus angular frequency $\omega$ and (b) the product of dielectric strength and \drts\ $\Delta\varepsilon\times {\sf g}(\log \tau)$ obtained for temperatures $322,~333,$ and $344\ \kelvin$. The symbols ($\circ$) and ($\diamond$) are the real and imaginary parts of the susceptibilities after subtraction of $\varepsilon'(\omega>\omega_\uparrow)$--they are the measured data from experiments. Solid (\full) lines represent the reconstructed resposes using the distribution functions ${\sf g}(\log \tau)$ for the real and imaginary part of the dielectric susceptibilities, respectively.}
\end{figure}

The analyses by means of the \drts\ for three temperatures, $322,~333,$ and $344\ \kelvin$, are displayed in Figure~\ref{fig:temp130}. In the figure, curves in the upper graphs (Figure~\ref{fig:temp130}a) are the real and imaginary parts of the susceptibilities, $\chi(\omega)=\varepsilon(\omega)-\varepsilon'(\omega>\omega_\uparrow)$, and the curves in the lower (Figure~\ref{fig:temp130}b) are the logarithm of the product of the total dielectric strength $\Delta\varepsilon$ and the obtained \drts\ ${\sf g}(\log \tau)$. The reconstructed dielectric susceptibilities are illustrated as solid (\full) % and dashed (\dashed) 
lines % in the upper graphs 
for real and imaginary parts. % respectively. 
The symbols in the figure are the measured susceptibilities $\chi$ after substruction of instantaneous permittivity $\varepsilon'(\omega>\omega_{\uparrow})$. The \drts\ is presented as histograms. Focusing on the three peaks obtained from the Monte Carlo analysis in the \drts\ spectra, Figure~\ref{fig:temp130}b, the most probable relaxation time behavior can be seen as stated by \textcite{AngellNgai}. Neglecting the weak relaxation at $\log\tau\sim-7.5$, and just considering the data in the angular frequency region $100\ \mega\reciprocal\second>\omega>20\ \giga\reciprocal\second$ -- one could assign a relaxation time using a curve-fitting procedure which would probably yield only one relaxation time (the superposition of the relaxations as observed in Figure~\ref{fig:temp130}b). As a remark, our analyses indicate that there exist a couple of relaxations taking place in \ppg.  Interestingly only one of the peaks move with temperature. The peaks at $10\ \giga\reciprocal\second$ are due to the preprocessing of the data since the spectra should start with a Debye relaxation~\cite{MacDonald1987}. However, this fast relaxation, which shows no temperature dependence, could also be treated as a boson peak, which seems to be  a general phenomenon of disordered molecular systems and not connected with a glass transition~\cite{GlassTrans}.
\begin{figure}[t]
  %\newpage
  \begin{center}
    {\includegraphics[]{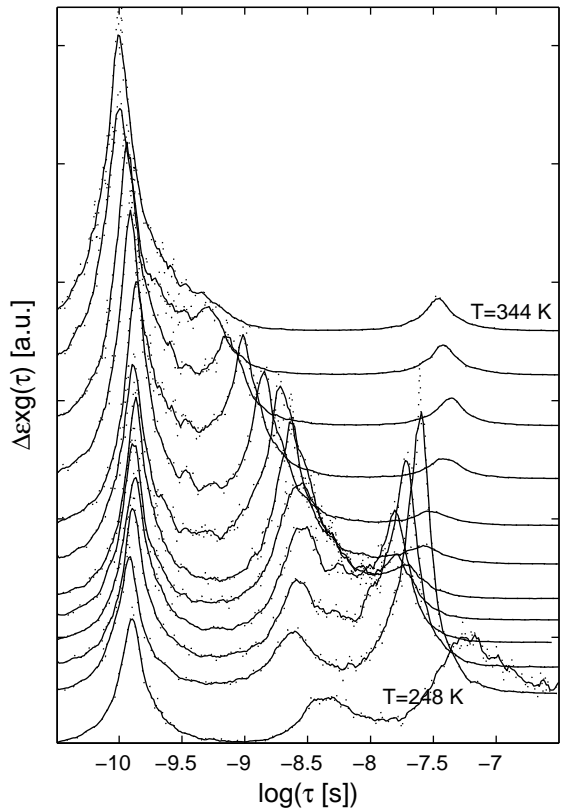}}
  \end{center}
  \caption{\label{fig:tauspec}The product of dielectric strength and the \drts\ [$\Delta\varepsilon\times {\sf g}(\tau)$] versus logarithm of relaxation time $\tau$ at different temperatures.}
\end{figure}

The time-temperature superposition states that by heating the material, the dielectric loss peak moves to higher frequencies. This movement is related to the thermally activated relaxation processes, that makes it possible to estimate the activation energy for the particular loss. In Figure~\ref{fig:tauspec} \drts\ versus relaxation time $\log \tau$ is plotted at different temperatures. The relaxations at very short times ($\log\tau\sim-10$) are due to the experimental conditions  since there exists a first relaxation process in the material with the application of the voltage \cite{MacDonald1987}. This relaxation could also be confused with the boson peak, which shows no temperature dependence as stated by \textcite{GlassTrans} and \textcite{Dionisio2000}. The amplitute of the high frequency relaxation increases as the temperature is increased due to the movement of the dielectric data to higher frequencies where they show a frequency-temperature superposition and interfere with a third peak not sensible to temperature. Besides, two peaks at low temperatures becomes one peak after $300\ \kelvin$. The two relaxations at low temperatures with slow relaxation times ($\log\tau\sim[-8.5\dots-7]$), do not mix with each other on the crossover region~\cite{GlassTrans}, which is clearly observed in the figure. The method applied to resolve the \drts\ gives a new insight into this region in glass-forming liquids and polymers. The amplitude of the slowest relaxation decreases just before the crossover temperature $T_c$ (approx. $275\ \kelvin$), however, the amplitude of the intermediate relaxation increases its amplitude as the temperature increases. The position of this peak starts to move to faster times as temperature values are over $275\ \kelvin$. 
\begin{figure}[t]
  %\newpage
  \centering
    {\includegraphics[]{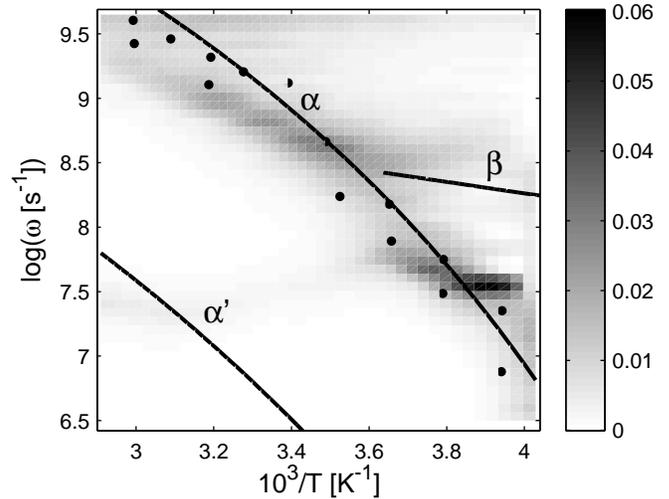}}    
  \caption{\label{fig:taubw}Relaxation-image for poly(propylene glycol). The solid lines (\full) represent the fitted curves of Ref.~\onlinecite{Furukawa} for $\alpha$- and $\beta$-relaxations, and of Ref.~\onlinecite{FKremerPPG} for the $\alpha'$-relaxation, respectively. The symbols ($\bullet$) represent the segmental relaxation for poly(propylene oxide) taken from Ref.~\onlinecite{Mijovic2002}.}
\end{figure}

In order to illustrate the temperature dependence of dielectric relaxation for \ppg, a different representation is shown in Figure~\ref{fig:taubw}.  In the image, the dark spots are the hills of the relaxations and white regions are plateaus where no relaxations are observed. The lines in the figure are taken from Arrhenius plots of previously published papers~\cite{Furukawa,FKremerPPG}. %The solid (\full), dashed (\dashed) and dotted (\dotted) lines are $\beta$-, $\alpha$ and $\alpha'$-relaxations, respectively. 
The solid line (\full) in the figure represents the fitted-curve to a VTF equation adapted from \textcite{Furukawa} for the $\alpha$-relaxation. Our temperature interval is just above the temperature region that \textcite{Furukawa} had in their experiments. At low temperatures the obtained \drts\ coincide with the fitted curve indicating that the main diagonal peak is the $\alpha$-relaxation. The solid line (\full) in the figure is an Arrhenius fit to the $\beta$-relaxation again from \textcite{Furukawa}. As stated by \textcite{Furukawa}, near $240\ \kelvin$ the $\alpha$- and $\beta$-relaxations separate, however, the \drts\ obtained demostrate that this separation or merging of the two relaxations occur around $275\ \kelvin$. The discrepancy could be due to the sample preparation and the position of the glass transition temperature which affects the cooperative processes in the sample. Moreover, the separation region in the relaxation-image is broad and the \drts\ analysis probes this region better than the conventional methods. The proximity of the $\beta$-relaxation peak induces a change in the $\alpha$-relaxation, which is clearly observed as a repulsion of two peaks in Figure~\ref{fig:tauspec}. It is interesting to see this effect as if the $\beta$-relaxation slows down the $\alpha$-relaxation as they approach  each other around $275\ \kelvin$. In addition the amplitutes of the two relaxations behave opposite to each such that the amplitude of the $\beta$-relaxation increases while the amplitude of the $\alpha$-relaxation decreases as they approach to each other.  This illustrates that the cooperative motion of the segments loose their character and some part of them join the local movement of the structural units, however, the activation energy of this new process is higher than the one for the local movements ($\beta$-relaxation). 

We should always keep in mind that the VTF equation uses the most probable relaxation time.  At high temperatures, the VTF equation is therefore not valid which is also expected after the merging of two relaxations. Although there is a good agreement between the data of previous investigations and ours, the peaks (dark spots) of the \drts\ is not coinciding exactly. The diagonal movement of the dominant peak in Figure~\ref{fig:tauspec} is significant and there are two horizontal spots that contribute to the diagonal main sequence. The data of \textcite{Mijovic2002}  are also illustrated in the figure with symbols ($\bullet$), which represent the segmental mode relaxation for for poly(propylene oxide), and it coincides with the $\alpha$ relaxation of \ppg\ as reported by \textcite{Furukawa}.

\begin{figure}[t]
  %\newpage
  \begin{center}
    {\includegraphics[]{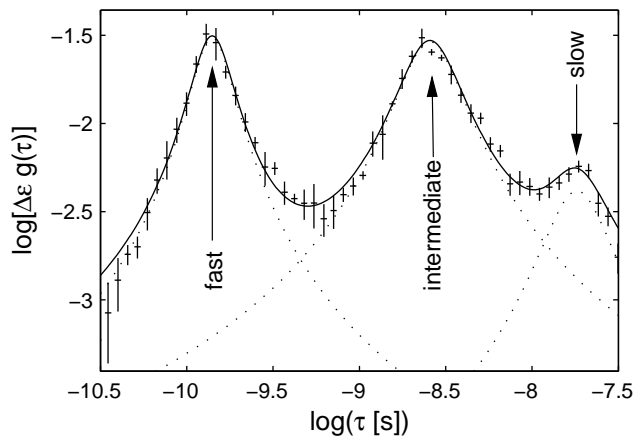}}    
  \end{center}
  \caption{\label{fig:spectrumfit} \drt\ at $299\ \kelvin$. The data obtained from the presented analysis is displayed with error bars. The dotted lines (\dotted) are the three Lorentz distributions fitted to the fast, intermediate and slow relaxations. The solid line (\full) shows the sum of these Lorentz distributions.}
\end{figure}
The obtained \drts\, as presented in Figure~\ref{fig:tauspec}, are analyzed by the Lorentz or the Cauchy distribution [Eq.~(\ref{eq:Cauchy})] to find the more exact positions and widths of the individual \drts, $\Delta\varepsilon\times{\sf g}(\tau) = {\sf C}(\tau;A,\bar{\tau},n)$. An example of this curve-fitting procedure is displayed in Figure \ref{fig:spectrumfit}, where the \drt\ results are shown with errorbars and the fitted Lorentz distributions are the solid lines and Table \ref{tabone} presents the parameters of the distributions at different temperatures. In the table relaxations in each temperature are separated into three different contributions, fast, intermediate and slow, respectively (also presented in Figure~\ref{fig:spectrumfit}). Similarly the data are fitted by Eq.~(\ref{eq:7}) and the fitting parameters are given in Table \ref{table2}. As mentioned previously, while discussing Figure~\ref{fig:eps0temp}, the curve fitting procedure with the Havriliak-Negami expression confirms that the dominant relaxation becomes Debye-like as the temperature approaches to $340\ \kelvin$. However, this procedure is neither probing into the crossover region or resolving the present two relaxations. %One can not be sure about the sides of the experimental window since there are not enough data points. 
In addition the Havriliak-Negami curve fitting procedure yields permittivity values at high frequencies $\varepsilon'(\omega>{\omega}_\uparrow)$ lower than one which is not physical for \ppg. Returning back to Table~\ref{tabone}, it is illustrative to see the change in the amplitude of the fast relaxation as the temperature is increased in the experiments. While comparing the amplitudes of the intermediate and slow relaxations for temperatures around the crossover temperature, we can decompose the overlapping dynamics of $\alpha$ and $\beta$ processes, using the probability theory. The presented analysis technique can provide a deeper insight of processes taking place in molecular systems such as dipole-dipole interaction.
\begin{table}[t]
  \caption{The Cauchy or Lorentz distribution [Eq.~(\ref{eq:Cauchy})] parameters for the \drts\ at different temperatures. \label{tabone}} 
%  \begin{tabular*}{3.3in}{@{\extracolsep{\fill}}{@{}*{1}{c}@{~}*{3}{c}@{~}*{3}{c}@{~}*{3}{c}}
  \begin{tabular*}{3.3in}{c r@{.}l r@{.}l c@{~~} r@{.}lc c @{~~} r@{.}l c r@{.}l}
    \toprule                              
                                %$A$&$B$&$C$&\m$D$&\m$E$&$F$&$G$\cr 
    &\centre{5}{Fast}&\centre{4}{Intermediate}&\centre{5}{Slow}\\
    \ns
    &\crule{5}&\crule{4}&\crule{5}\\
    $T$&\centre{2}{$A$}&\centre{2}{$\log\bar\tau$}&$n$&\centre{2}{$A$}&$\log\bar\tau$&$n$&\centre{2}{$A$}&$\log\bar\tau$&\centre{2}{$n$}\\ 
    $[\kelvin]$&\centre{2}{\footnotesize $[10^{-3}]$}&\centre{2}{}&$$&\centre{2}{\footnotesize $[10^{-3}]$}&$$&$$&\centre{2}{\footnotesize $[10^{-3}]$}&$$&\centre{2}{$$}\\ 
    \colrule
    248&7&3 &$-9$&$9$&9.5& 3&5 &$-8.4$&7.5  &22&  &$-7.1$&3&3\\
    260&8&3 &$-9$&$9$&8.7& 6&2 &$-8.6$&4.8  &17&  &$-7.6$&8&5\\
    266&9&1 &$-9$&$9$&8.9& 9&7 &$-8.5$&4.3 &14&  &$-7.8$&7&7\\
    272&9&9 &$-9$&$9$&8.9& 14&   &$-8.5$&4.3 &8&4&$-7.8$&7&7\\
    277&10&   &$-9$&$9$&8.7& 18&   &$-8.5$&4.3 &3&7&$-7.8$&7&7\\
                                %\mr
    282&12&   &$-9$&$9$&8.5& 17&   &$-8.6$&5.2 &1&8&$-7.7$&8&0\\
    290&13&   &$-9$&$9$&7.2& 15&   &$-8.7$&5.8 &0&8&$-7.6$&10&\\
    299&16&   &$-9$&$8$&7.6& 12&   &$-8.9$&6.2 &0&7&$-7.5$&10&\\
    310&18&   &$-9$&$9$&7.0& 10&   &$-9.0$&6.6 &1&1&$-7.4$&9&8\\
    322&20&   &$-9$&$9$&7.0&7&0  &$-9.2$&6.3 &1&5&$-7.4$&10&\\
    334&20&   &$-10$&$ $&6.7&5&8  &$-9.4$&5.5 &1&7&$-7.4$&9&7\\
    344&21&   &$-10$&$ $&7.0&3&8  &$-9.5$&4.8 &1&9&$-7.4$&9&3\\
                                %266 &0 &0.75&$-57.2$&\m---   &---  &---\\ 
                                %272 &60  &0.60&$-48.1$&$-0.29$ &41   &15\\ 
    \botrule
  \end{tabular*}
                                %\end{indented}
\end{table}

\begin{figure}[t]
  %\newpage
  \begin{center}
    {\includegraphics[]{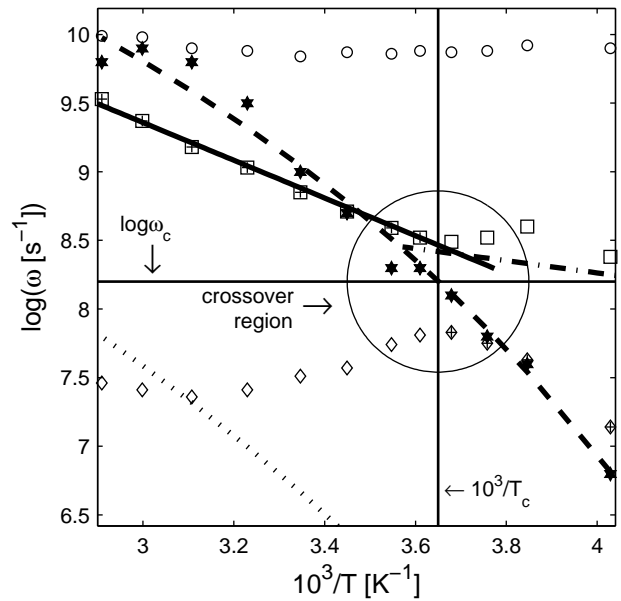}}    
  \end{center}
  \caption{\label{fig:arrhenius} Arrhenius plot of \ppg. Symbols show the result of the present study. Open symbols [($\circ$)--fast relaxation, ($\Box$)--intermediate relaxation, ($\diamond$) slow relaxation] are relaxation rates obtained from \drt\ analyses (Table~\ref{tabone}) and filled symbols ($\star$) are the curve fitting results of the employed Havriliak-Negami empirical formula (Table~\ref{table2}).  The points represented by ($+$) symbols are the relaxations with large amplitudes in Table~\ref{tabone} ($A>5\times10^{-3}$). The solid (\full) line is a straight line fitted to the intermediate relaxation at high temperatures ($T>275\ \kelvin$). The dashed (\dashedd) and chain (\chain) lines are the VTF and Arrhenius fits from literature~\cite{Furukawa} for the $\alpha$ and $\beta$ relaxations, respectively. The dotted (\dotted) line is the VTF fit for the $\alpha'$-relaxation~\cite{FKremerPPG}. The figure is divided into four sectors taking the crossover point $[10^3/T_c,\ \log\omega_c]$ as center.}
\end{figure}

Last but not least we display in Figure~\ref{fig:arrhenius} the Arrhenius plot of \ppg. In the figure the logarithm of inverse relaxation times ($\tau^{-1}=\omega$) are presented. The results obtained from both the Havriliak-Negami empirical formula and the \drt\ method are used. The open symbols ($\circ$, $\Box$ and $\diamond$) show the \drt\ analyses (Table~\ref{tabone}). The filled symbols ($\star$) are the relaxation times obtained from the curve-fitting of the empirical formula (Table~\ref{table2}). To give some more insight, data from the literature are also added; the dashed (\dashedd) and chain (\chain) lines are from ~\textcite{Furukawa}, and the dotted line (\dotted) is from~\textcite{FKremerPPG}. The graph is divided into four sectors, which are used in the discussion below. The origin of the sectors is the crossover temperature $T_c$ and the corresponding relaxation rate ($\log\omega_c$). In sector 1, there is only the trace of the $\beta$ relaxation and the \drt\ analysis yields relaxation rates little higher than stated in the literature~\cite{Furukawa}. In sector 2, there is a couple of points that should be mentioned. First, the curve-fitting method and \drt\ result in different relaxation time constants. The \drt\ result in a linear Arrhenius relation, $\log\omega=a+b/(T)$ with $a=13.47$ and $b=-1374$, while the curve-fitting method show a good agreement with the VTF equation reported in literature~\cite{Furukawa}. One should keep in mind that our experimental interval is just outside of those in \textcite{Furukawa}, and the line drawn is an extrapolation of their results. In sector 3 there is only the weak slow relaxation, which is not as significant as the others in the \drt\ spectra. The relaxation could be associated with the $\alpha'$-relaxation. However, we should keep in mind that the relaxation has a weak amplitude and data is just at the border of our experimental window.  Lastly in sector 4 the $\alpha$ relaxation is present and it vanishes as it approaches the crossover temperature $T_c$. Although the Havriliak-Negami empirical formula produce valuable results, it is not sufficient to give insight into the crossover region as the \drt\ analysis. It should be noted that the behavior of three relaxations shown in Figure~\ref{fig:arrhenius} are very similar to those illustrated for polyethyl methacrylate in page 29 of \textcite{GlassTrans} also in Ref. \onlinecite{Schroter1998}, for poly(alkyl methacrylate)s of \textcite{Beiner2001} and for other glass-forming systems \cite{Gomez2001,Wang2002,AngellNgai,Huang2002}.

\begin{table}[t]
  \newpage
  \caption{Fitting parameters obtained by applying the Havriliak-Negami empirical formula in Eq.~(\ref{eq:7}).\label{table2}}
  \begin{tabular*}{3.3in}{@{\extracolsep{\fill}}cccccc}
    \toprule                         
    T[K] &$\varepsilon'$&$\Delta\varepsilon$&$\log\tau$&$\alpha$&$\beta$\\
    \colrule
    248& 1.0 &   2.35 & 6.0  & 0.58  & 0.91\\
    260& 1.0 &   1.76 & 6.8  & 0.77  &  0.72\\ 
    266& 1.0 &   1.67 & 7.0  & 0.82  &  0.66\\ 
    272& 1.0 &   1.59 & 7.3  & 0.80  & 0.75\\ 
    277& 1.0 &   1.49 & 7.5  & 0.80  & 0.81\\ 
    282& 0.9 &   1.58 & 7.5  &   0.90  & 0.50\\ 
    290& 1.0 &   1.11 & 7.9  & 0.86  & 0.96\\ 
    299& 0.9  &   1.18 & 8.2  & 0.82  & 0.97\\ 
    310& 1.0  &   1.79 & 8.7  & 0.74  & 0.96\\ 
    322& 0.9  &   1.84 & 9.0  & 0.75  & 0.97\\ 
    334& 1.0 &   1.63 & 9.1  & 0.83  & 1.00\\ 
    344& 1.4 &   1.25 & 9.0  & 1.00  & 0.98\\ 
    \botrule
  \end{tabular*}
\end{table}

%\begin{figure}[t]
%  \begin{center}
%    {\includegraphics{responcefunc.eps}}    
%  \end{center}
%  \caption{\label{fig:response}Time-domain response function for \ppg\ at different temperatures.}
%\end{figure}

%\begin{figure}[t]
%  \begin{center}
%    \includegraphics[width=4in]{rcm3d.eps}    
%  \end{center}
%  \caption{\label{fig:tau3d}Surface plot of relaxation map for poly(propylene glycol).}
%\end{figure}

\section{Conclusions}

A numerical method is applied to dielectric data of \ppg\ to extruct the \drts. To demonstrate the potential of the method, the \drt\ of the known distributions, delta function and box, were obtained from generated data. Later the same procedure was applied to experimental data of \ppg. The method was able to find a unique \drts\ for given dielectric data.  It was illustrated that for molecular systems we were not limited to analyze average or most probable relaxation times. However, new questions arise for the origin of the obtained \drts.

\ppg\ showed two significant \drts\ which are visible between $260-277\ \kelvin$. At the higher temperatures there was only one visible peak in the \drts\ spectra. The most significant relaxation below $275\ \kelvin$ was associated with the $\alpha$ relaxation. The second significant relaxation on the same temperature region was associated to the $\beta$-relaxation. It was observed that the $\alpha$- and $\beta$-relaxation separation (as well as merging) (crossover region) was around $275\ \kelvin$. As the temperature increased above the crossover temperature, only one relaxation was significant, and it had a linear Arrhenius relation. It appeared as if the $\beta$-relaxation slowed-down the $\alpha$-relaxation which proved the strength of the applied analysis method  over the traditional curve-fitting procedures. Comparison of the \drt\ approach to those of the curve-fitting one have especially pointed out the potential of the \drt\ technique in the case of the crossover region. The process at temperatures above the crossover temperature was not actually a continuation of the $\alpha$ relaxation, which shows a VTF type temperature dependence, but it was clearly a continuation of the $\beta$-relaxation with a higher activation energy. The comparison of the Arrhenius- or relaxation-image with data from literature verified that the method was successful for low temperatures, $\alpha$- and $\beta$-relaxations. Although, the fast relaxation resolved from the dielectric data coincides with the boson peak speculation in disorder molecular systems~\cite{GlassTrans}, other indications are needed to support the results.

The \drts\ illustrated the importance of interaction between different processes  %(dipole-dipole interactions) 
in molecular systems such that as the merging of the $\alpha$- and $\beta$-relaxations. Their interaction could be observed by the presented method. The cooperative motion of the chain units ($\alpha$ process) creates space for the local motion ($\beta$-process). As the sample is heated only the local motion of polymer units survives as a detectable relaxation process, and it shows an Arrhenius temperature dependence. It is important to continue dielectric measurements on \ppg\ samples with smaller temperature intervals and a broader frequency window to better understand the nature of molecular processes taking place. 

Finally, we in this article gave our attention to one of the ignored subjects, the \drts\, in condensed-matter physics. It is illustrated through out the text that it can be used to obtain valuable information from the dielectric data, which is not possible otherwise by classical dielectric data analyses approaches whose main assumption is using the average or most probable relaxation time for a given dielectric loss peak. As a metaphor, it is important to state that in early days of astronomy well-known pioneers in the field used brightness of stars which were sum of all light from stellar-objects. Later researchers used the optical spectra to obtain abundances of different elements, which in return supplied valuable information for gaining a better understanding of stellar-objects. Well, traditional methods to analyze dielectric data only considers the shape of the dielectric data, and do not extract what is underneath. We, therefore, think that the applied numerical method acts in a similar way to the optical spectroscopy for astronomy, and resolves different relaxations,  {\em ``spectral-lines''}, hidden in the dielectric data. The next step is then to relate these individual relaxations to structure and abundances of molecular species in the material, and better understand whether the lines are homogeneously or inhomogeneously broadened. 

\section*{Acknowledgement}
One of the authors (ET) thanks Prof. Reimund Gerhard-Multhaupt for constructive discussion on temperature independent dipolar movements in the spectra and motivation for adapting the \cnlsq\ preprocessing algorithm for the high frequency permittivity estimate.

%\newpage
%\section*{References}
\bibliography{../../newref.bib}
\bibliographystyle{apsrev}

\end{document}